\documentstyle[12pt]{article}
\newcommand{\be}{\begin{equation}}
\newcommand{\ee}{\end{equation}}
\newcommand{\bea}{\begin{eqnarray}}
\newcommand{\eea}{\end{eqnarray}}
\newcommand{\al}{\alpha}

\newcommand{\gm}{\gamma}
\newcommand{\Gm}{\Gamma}
\newcommand{\dl}{\delta}
\newcommand{\Dl}{\Delta}
\newcommand{\eps}{\epsilon}

\newcommand{\lm}{\lambda}

\newcommand{\om}{\omega}

\newcommand{\pa}{\partial}
\newcommand{\dd}{\mbox{d}}

\newcommand{\ux}{\underline{x}}

\newcommand{\uq}{\underline{q}}
\newcommand{\uk}{\underline{k}}
\newcommand{\um}{\underline{m}}

\newcommand{\ual}{\underline{\alpha}}
\newcommand{\nn}{\nonumber}
\newcommand{\SH}{\hat{S}}
\newcommand{\smi}{\setminus}
\newcommand{\pasl}{\not\! \partial}
\newcommand{\pesl}{\not\! p}
\newcommand{\qsl}{\not\! q}
\newcommand{\ksl}{\not\! k}

\begin{document}
\parindent=1.5pc
\begin{titlepage}
\begin{center}
{{\bf
Gauge-Invariant Differential Renormalization: Abelian
Case\footnote{Supported by INTAS, grant 94--4666, and by 
the Russian Foundation for Basic Research, 
project 96--01--00726.}}\\
\vglue 5pt
\vglue 1.0cm
{ {\large V.A. Smirnov}\footnote{E-mail: smirnov@theory.npi.msu.su } }\\
\baselineskip=14pt
\vspace{2mm}
{\it Nuclear Physics Institute of Moscow State University}\\
{\it Moscow 119899, Russia  }\\
\vglue 0.8cm
{Abstract}}
\end{center}
\vglue 0.3cm
{\rightskip=3pc
 \leftskip=3pc
\noindent
A new version of differential renormalization is presented.
It is based on pulling out certain differential operators and
introducing a logarithmic dependence into diagrams.
It can be defined either in coordinate or momentum space,
the latter being more flexible for treating tadpoles and diagrams
where insertion of counterterms generates tadpoles.
Within this version, gauge invariance
is automatically preserved to all orders in Abelian case.
Since differential renormalization is a strictly four-dimensional
renormalization scheme
it looks preferable for application in each situation when dimensional
renormalization meets difficulties, especially, in theories with
chiral and super symmetries.
The calculation of the ABJ triangle anomaly is given as an example
to demonstrate simplicity of calculations within
the presented version of differential renormalization.
\vglue 0.8cm}
\end{titlepage}

\section{Introduction}

Various versions of differential renormalization [1--7] explicitly describe
how a product of propagators corresponding to a given graph and
considered in coordinate space can be defined as a distribution
on the whole space of test functions, starting from the subspace
of test functions which vanish in a vicinity of points where the
coordinates coincide. Differential renormalization provides
a strictly four-dimensional renormalization scheme useful for
applications \cite{appl1,appl2}.
It is based on `pulling out' some
differential operators from the initial unrenormalized diagram;
these are either $\Box = \pa_{\al} \pa_{\al}$ [1--4,7], or
$\hat{S} = \frac{1}{2} \frac{\pa}{\pa x_{\al}} x_{\al}$
\cite{SZ,S2}. The latter version looks more natural for generalization
to the case with non-zero masses when one uses the operator
\be
\hat{S} = \frac{1}{2} \sum_{i} \frac{\pa}{\pa u_{i}} u_{i}
- \frac{1}{2} \sum_{l}
m_{l} \frac{\pa}{\pa m_{l}} ,
\label{SXM}
\ee
where $u_i$ are independent coordinate differences.

The purpose of this paper is to present a new version of differential
renormalization which automatically preserves gauge symmetry at least
in the Abelian case. To do this we shall modify, in the next section,
the version of ref.~\cite{SZ,S2} based on operator (\ref{SXM}).
The renormalization is defined recursively, for every graph, provided
it was defined for graphs and reduced graphs with a smaller number
of loops. It reduces to pulling out the operator $\SH$ and
inserting some logarithm into the given Feynman amplitude.
The main idea of the new version of differential renormalization is
to `spoil' the diagram by the logarithms in a minimal way, in the
sense that a minimal number of propagators is spoiled by the
logarithms.

In Section~3 we shall present a gauge-invariant differential
renormalization scheme
for QED. The underlying idea is simple: to insert the logarithms
into the photon lines. This enables us to perform, for renormalized
Feynman integrals, the same manipulations for establishing the gauge
invariance as for unrenormalized ones.
Thus gauge invariance is maintained automatically, to all
orders.\footnote{In \cite{KMZ} a recipe of differential renormalization
formulated in momentum space and applicable for diagrams with
simple topology of divergences was applied to prove relations
relevant to Ward identities and corresponding to some partial classes
of diagrams that contribute to Green functions.}
One-loop polarization of vacuum requires special treatment
when an operator with explicitly transverse structure is pulled out
before renormalization.

In Section~4, by translating renormalization prescription into the
language of momentum space, we shall present general recipes applicable to
arbitrary diagrams. In particular, we shall treat, in a more simple way
(as compared to ref.~\cite{S2}), renormalization of tadpoles and diagrams
where insertion of counterterms for subgraphs generates tadpoles (e.g.
the sun-set diagram). Furthermore, the calculation of the triangle ABJ
anomaly is presented as an example of the application of differential
renormalization.

Finally, Section~5 contains discussion of the obtained results.

\section{Differential renormalization \newline in coordinate space}

\subsection{Renormalization prescriptions}

A Feynman amplitude is defined in coordinate space through
products of propagators
\be
\Pi_{\Gm} (x_1, \ldots , x_N) \equiv \Pi_{\Gm} (\ux) =
\prod_{l} {\textstyle{1\over i}} G_l (x_{\pi_+(l)} - x_{\pi_-(l)}),
\label{F0b}
\ee
taken over lines of a given graph $\Gm$.
Here $\pi_{\pm}(l)$ are respectively beginning and the end of a line $l$,
and
\be
G_l (x) = \frac{i m_l}{4\pi^2} \;
P_l (-i \pa / \pa x, m) \;
\frac{K_1 (m_l \sqrt{-x^2+i0})}{\sqrt{-x^2+i0}}
\equiv P_l (-i \pa / \pa x, m) G(x)
\label{PROP}
\ee
is a propagator, with $P_l$ polynomial and $K_1$ a modified Bessel
function.

The strategy of differential renormalization is an explicit
realization of the extension of functional (\ref{F0b}), from
a subspace of test functions to the whole space
(see details in refs.~\cite{SZ,S2}).
Instead of renormalization prescriptions of \cite{SZ,S2},
let us now define renormalization of Feynman amplitudes by the following
recursive procedure. Note that we need to renormalize not only
Feynman amplitudes themselves but as well Feynman amplitudes
$\Pi^{j,l}_{\Gm} (\ux)$
`spoiled' in some way by additional logarithmical dependence.
This notation implies that the $j$-th power of some logarithm
is introduced in a certain way into some fixed $l$-th line:
\be
\Pi^{j,l}_{\Gm} (\ux) = {\textstyle{1\over i}}
G^{(j)}_l (x_{\pi_+(l)} - x_{\pi_-(l)})
\prod_{l'\neq l} {\textstyle{1\over i}}
G_l (x_{\pi_+(l')} - x_{\pi_-(l')}) .
\label{F00b}
\ee
For example, one can substitute $G_l (x)$ (corresponding to this very line)
by
\be
G^{(j)}_l (x) = \ln^j \mu^2 x^2 G_l (x) .
\label{Gjlx}
\ee
However this is not the simplest version suitable for applications.
A better variant is based on multiplication by a logarithm
in momentum space:
\be
G^{(j)}_l (x) = \frac{1}{(2\pi)^4} \int \dd^4 q \;
(-1)^j \ln^j\left( H(q) / \mu^2\right) \tilde{G}_l (q) ,
\label{Gjlq}
\ee
where the function $H$ can be chosen as $-q^2-i0$ or
$m_l^2-q^2-i0$. The latter version generally looks much more
simple from the point of view of the calculational simplicity.
In all these situations, the logarithmically spoiled propagator
satisfies
\be
\hat{D} G^{(j)}_l (x) = j G^{(j-1)}_l (x) - (1+a_l) G^{(j)}_l (x)
\label{PROPj}
\ee
with
\be
\hat{D} = \frac{1}{2} x \frac{\pa}{\pa x}
- \frac{1}{2} m_l \frac{\pa}{\pa m_{l}} \equiv \hat{S} -2
\label{DXM}
\ee
and $a_l$ degree of the polynomial $P_l$ in (\ref{PROP}).

Suppose that we know how to renormalize all the Feynman amplitudes
$\Pi^{j',l}_{\gm} (\ux)$ and $\Pi^{j',l}_{\Gm/\gm} (\ux)$ corresponding
to reduced graphs $\Gm/\gm$ and proper subgraphs $\gm \subset \Gm$
of the given graph $\Gm$. (These subgraphs and reduced graphs
have smaller number of loops, $h$. Recall that the reduced graph
$\Gm/\gm$ is obtained from $\Gm$ by reducing each connectivity
component of $\gm$ to a point.) Here $l$ is a fixed line
of $\Gm$, and $j'$ is arbitrary integer.
This means that we know all counterterms contributing to the
$R$-operation (i.e. renormalization at the diagrammatical level)
which acts on the corresponding Feynman amplitudes:
\bea
R\Pi^{j',l}_{\gm} = \sum_{\gm_1, \ldots, \gm_j}
\Delta(\gamma_1) \ldots \Delta(\gamma_j) \Pi^{j',l}_{\gm} \nonumber \\
\equiv R'\Pi^{j',l}_{\gm} + \Delta(\gm) \Pi^{j',l}_{\gm} .
\label{R1PI}
\eea
where $\gm$ stands  either for a subgraph or reduced graph,
$\Delta$ is the corresponding counterterm operation, and
the sum is over all sets $\{\gm_1, \ldots, \gm_j\}$
of disjoint divergent 1PI subgraphs of $\gm$,
with $\Dl(\emptyset)=1$. The operation $R'$ is called
incomplete $R$-operation.

In this and the following section we shall
consider diagrams without tadpoles and such that
insertion of counterterms for subgraphs does not generate tadpoles.
Note that this is sufficient for QED.

Let us now define renormalization of the Feynman amplitude
$\Pi^{j,l}_{\Gm}$ as:
\be
R\,  \Pi^{j,l}_{\Gm}  = \frac{1}{j+1}
\left( \hat{S} + \om /2 \right)\left (1-M^{\om-1}\right)
\; R'\, \Pi^{j+1,l}_{\Gm} -
\frac{1}{j+1} \sum_{\gm \subset \Gm:\; l\overline{\in} \gm} R
\, {\cal C}_{\gm} \Pi^{j+1,l}_{\Gm},
\label{F17}
\ee
Here $R'$ is incomplete $R$-operation given by (\ref{R1PI}),
and $\om=4h-2L+a$ is the degree of divergence (with $L$ the number
of lines and $a$ total degree of the polynomials $P_l$ in (\ref{PROP})).
The sum in the second term of the right-hand side of (\ref{F17})
runs over all 1PI proper subgraphs $\gm$ of $\Gm$ that do not include
the line $l$.
The operator $M^{r}$ performs Taylor expansion of order $r$ in masses
and external momenta
of the graph $\Gm$.

Furthermore,
\be
\Dl(\Gm) \Pi^{j,l}_{\Gm} = R \Pi^{j,l}_{\Gm} - R'\, \Pi^{j,l}_{\Gm}
\label{Dl}.
\ee
Finally, the operations ${\cal C}_{\gm}$
are determined from equations
\bea
\left( \hat{S} + \om /2 \right)
R\, \Pi_{\Gm} = \sum_{\gm \subseteq \Gm}
{\cal C}_{\gm} R \Pi_{\Gm} \equiv \sum_{\gm \subset \Gm}
R {\cal C}_{\gm} \Pi_{\Gm } + {\cal C}_{\Gm} \Pi_{\Gm },
\label{F18}
\eea
where the operation
${\cal C}_{\gm}$ inserts a polynomial ${\cal P}_{\gm}$
of degree $\om(\gm)$ in masses
of $\gm$ and its external momenta into the reduced vertex of the graph
$\Gm /\gm$.
Symbolically we write
\be
{\cal C}_{\gm} \Pi_{\Gm } = \Pi_{\Gm /\gm} \circ {\cal P}_{\gm}
\label{CG}
\ee
where $\circ$ denotes the insertion operation. In the language of
coordinate space,
\be
{\cal C}_{\Gm} \Pi_{\Gm } (\ux) =
{\cal P}_{\Gm} (\pa /\pa x_i) \prod_{i \in \Gm, i\neq i_0}
\dl^{(4)}(x_i-x_{i_0} ) ,
\label{CGC}
\ee
where $i_0$ is a fixed vertex. The polynomial for the graph ${\cal P}_{\Gm}$
is expressed  from the difference of the left-hand side of
(\ref{F18}) and the sum in the right-hand side.

\subsection{Comments and examples}

{\bf 1.} A proof of the fact that relations (\ref{F17})--(\ref{F18})
define a correct $R$-operation can be obtained by a trivial
modification of the arguments presented in ref.~\cite{S2}.
In particular, it is proved that in the coordinate space with deleted origin
(i.e. when at least one of the coordinate difference variables $x_i-x_j$
is non-zero) the incompletely renormalized Feynman amplitude
$R'\, \Pi^{j,l}_{\Gm}$ can be written as the
right-hand side of eq.~(\ref{F17}). Then this equation enables us
to extend the initial functional to the whole space of test functions
so that, by definition, the right-hand side of eq.~(\ref{F17})
determines  the renormalized quantity $R\, \Pi^{j,l}_{\Gm}$.
Indeed, the first term there does not have divergences
because all subdivergences are removed by $R'$. As to the overall
divergence, it is removed by the product of the operators
$\left( \hat{S} + \om /2 \right)\left (1-M^{\om-1}\right)$.
The preliminary Taylor subtractions result in producing a
polynomial in momenta and masses of degree $\om$ and thereby
reduce the degree of divergence from $\om$ to zero. 
After commutation of the operator
$\left( \hat{S} + \om /2 \right)$ with this polynomial, the resulting
operator $\hat{S}$ acts as in logarithmically divergent case
(see details in \cite{SZ,S2}) by removing the divergence by multiplication
of the first degree monomials contained in operator (\ref{SXM}).

The other terms in the right-hand side of eq.~(\ref{F17}) are
manifestly renormalized Feynman amplitudes corresponding to
reduced subgraphs (with smaller numbers of loops).

{\bf 2. Examples.} ({\em i}) If a diagram is primitively divergent
(i.e. does not involve subdivergences), with degree of divergence $\om$, then
the above prescriptions give
\be
R \Pi_{\Gm}  =
\left( \hat{S} + \om /2 \right)\left (1-M^{\om-1}\right)
\Pi^{1,l}_{\Gm},
\label{prd}
\ee
where $l$ is an arbitrary line of $\Gm$.
Instead of $\Pi^{1,l}_{\Gm}$ one can also use a linear combination
$\sum \xi_l \Pi^{1,l}_{\Gm}$, with $\sum \xi_l =1$ (see e.g.
Sect.~4.2 where the triangle anomaly is calculated).

({\em ii}) To spoil the minimal number of the propagators involved
it is natural to choose, as the line $l$, a line that is
chosen for renormalization of some maximal subgraph of the given graph.
Let $\Gm$ be logarithmically divergent and let it involve only
one proper divergent subgraph $\gm$, with $\om(\gm)=0$. Then, using
some $l\in\gm$, one has
\be
R \Pi_{\Gm}  =
\frac{1}{2} \SH \left( \Pi_{\Gm\smi\gm} \SH \Pi^{2,l}_{\Gm} \right).
\label{2ds}
\ee

A generalization of this formula for the case when all
the divergent subgraphs of the diagram form a nest
$\gm_1 \subset \gm_2 \subset \ldots \subset \gm_r \equiv \Gm $,
with $\om(\gm_i)=0$, and the initial Feynman amplitude itself
involves the $j$-th power of the logarithm, looks like
\be
R \Pi_{\Gm}  =
\frac{(j-1)!}{(j+r)!} \SH \left( \Pi_{\Gm\smi\gm_{r-1}}
\SH \left( \Pi_{\Gm\smi\gm_{r-2}}
\ldots \SH \Pi^{j+r,l}_{\Gm} \right)\right) ,
\label{rds}
\ee
where $l\in\gm_1$.

({\em iii}) Let $\Gm$ be logarithmically divergent and let it involve
only two disjoint divergent subgraphs with $\om(\gm_i)=0$.
Let $l_i\in\gm_i, \; i=1,2$.
Then one can define
\be
R \Pi_{\Gm}  =
\SH \Pi_{\Gm\smi (\gm_1 \cup \gm_2)}
\left( \frac{1}{2} \SH \Pi^{2,l_1}_{\gm_1} \right)
\left( \SH \Pi^{1,l_2}_{\gm_2} \right)
- R  \left( {\cal C}_{\gm_2}  \Pi^{1,l_1}_{\Gm}\right) ,
\label{3ds}
\ee
where the operation ${\cal C}_{\gm_2}$ is defined by
\[ {\cal C}_{\gm_2} \Pi_{\gm_2} = \SH \Pi_{\gm_2}
\equiv c_{\gm_2} \prod_{i \in \Gm, i\neq i_0} \dl^{(4)}(x_i-x_{i_0} ). \]

({\em iv}) Let $\Gm$ possess the following family of
logarithmically divergent subgraphs: $\gm_1, \gm_2,
\gm_{12} \equiv \gm_1 \cap \gm_2$ and $\Gm$ itself. Let $l\in \gm_{12}$.
Then
\be
R \Pi_{\Gm}  = \SH R' \Pi^{1,l}_{\Gm}, \;\;
R'=1 + \Dl(\gm_1) + \Dl(\gm_2) + \Dl(\gm_{12}).
\label{ol1}
\ee
Here $\Dl(\gm_i)=R(\gm_i)-R'(\gm_i), \Dl(\gm_{12}) = R(\gm_{12})-1$ 
are found from (\ref{2ds}) and (\ref{3ds}) for $j=1$.

({\em v}) Let $\Gm$ be as in the previous example and $\gm_{12}\equiv l$
(which is not of course divergent).
Let us keep in mind the 2-loop photon exchange diagram of QED
contributing to the vacuum polarization. We have $\om(\Gm)=2$
and $\om(\gm_i)=1$ for one-loop vertex subgraphs. Then
\bea
R \Pi_{\Gm}  = (\SH+1) (1-M^1) R' \Pi^{1,l}_{\Gm},
\label{ol2} \\
R'=1 + \Dl(\gm_1) + \Dl(\gm_2), \;\; \Dl(\gm_i)=R(\gm_i)-1, \\
R \Pi_{\gm_i} = (\SH+1/2) (1-M^0) \Pi^{1,l}_{\gm_i}, \\
R \Pi^{1,l}_{\gm_i} = \frac{1}{2} (\SH+1/2) (1-M^0) \Pi^{2,l}_{\gm_i}.
\eea

{\bf 3.} We use here the operator
$\left( \hat{S} + \om /2 \right)\left (1-M^{\om-1}\right)$
instead of a differential operator of order $\om+1$ of
the previous version \cite{SZ,S2}.
Although this mixture of coordinate and momentum space operators
might seem strange we insist that this version
of differential renormalization is more simple that previous ones and
this preliminary momentum subtraction is natural.
In Section~4 this version will be translated into momentum space
which happens to be appropriate for treating arbitrary diagrams.
To apply the operator $(1-M^{\om-1})$
one can turn to momentum space, perform Taylor
expansion and then come back to coordinate space.
However this momentum subtraction is as well easily described in coordinate
space. In particular, in the massless case, we have the following
prescription for the action of the functional
$(1-M_q^{\om-1}) \Pi(x)$ on a test function $\phi(x)$:
\be
\left(  (1-M_q^{\om-1}) \Pi(x), \; \phi(x) \right) =
\int \dd x \Pi(x) (1-M_x^{\om-1}) \phi(x) ,
\ee
where $M_x^{\om-1}$ performs Taylor expansion in $x$.

{\bf 4.} It is also possible to `logarithmically spoil' the diagrams in the
language of the $\al$-representation, by inserting factors like
$\ln^j \mu^{2h} D(\ual)$ into the integrand of the
$\al$-representation (here $D(\ual)$ is a standard homogeneous
function). However this possibility looks disadvantageous because
of the lack of control of manipulations that are relevant
for establishing desired symmetry. Note that
momentum space modification (\ref{Gjlq}) of the propagator
is easily described in the $\al$-representation as insertion
of $\ln \mu'^2 \al_l$ (with $\mu'$ proportional to $\mu$ in
(\ref{Gjlq})).

{\bf 5.} For the products of propagators in coordinate space
all the vertex are considered as external. Feynman amplitudes
are obtained from the products $\Pi_{\Gm}$
by integrating over coordinates associated with internal
vertices:
\be
F_{\Gm}(x_1, \ldots, x_n) =
\int \dd x_{n+1} \ldots \dd x_N \Pi_{\Gm} (x_1, \ldots, x_N) .
\label{FPI} \ee
When writing down renormalization prescriptions (\ref{F17})--(\ref{F18})
for Feynman amplitudes (\ref{FPI})
one obtains similar formulae where the operator $\SH$ is now given by
the sum over external coordinates.

{\bf 6.} The coefficients polynomials ${\cal P}_{\gm}$ play the role
of contributions of simple poles to counterterms within  dimensional
renormalization. They are related with renormalization group functions
by the same formulae as in the case of the previous version of
differential renormalization --- see \cite{S2}.

\section{Gauge-invariant differential renormalization of QED}

As is well-known the Ward identities in QED, e.g.
\be
q_{\al} \Gm^{\al} (p,q,p+q) = \Sigma(p+q) - \Sigma(p) ,
\label{WI}
\ee
that connects the vertex and self-energy Green functions
are proved using standard manipulations based on the following identity
involving the electron propagator:
\be
\frac{\pa}{\pa \xi_{\al}} \left[ S(x_2-\xi) \gm_\al S(\xi -x_1)
\right]
= \frac{1}{i} [ \dl(x_2 -\xi) - \dl(\xi -x_1) ] S(x_2 -x_1),
\label{id}
\ee
which are based on the equation of motion for the free electron propagator
$S(x)\equiv (m + i \pasl ) G(x)$, namely $(m - i \pasl ) S(x) = \dl(x)$.

In fact one uses (see, e.g., \cite{BS}) a natural one-to-one
correspondence between diagrams that contribute to these Green functions
(vertex diagrams are obtained by insertion a new triple vertex into
lines of the electron path between external electron lines).
The main problem in establishing the Ward identity
is to prove that these manipulations are also
valid for the {\em renormalized} Feynman amplitudes.
Let us now use the structure of our renormalization
procedure which reduces to pulling out differential operators and
spoiling the Feynman amplitudes by logarithms. Note that commutation
relations of the differential operators with monomials in external
momenta (i.e. derivatives in coordinates) are very simple:
\be
P^r \hat{S} = (\hat{S} + r/2) (1-M^{r-1}) P^r ,
\label{CR}
\ee
where $P^r$ is such a monomial of degree $r$ (the value $r=1$ is relevant
for (\ref{WI})).

Therefore, the problem is not to spoil identities (\ref{id})
by the inserted logarithms. Since these identities
are connected with electron lines a natural and simple
solution of this problem is
to introduce the logarithms only into photon lines.
Then the proof of (\ref{WI}) is performed by induction (as
the renormalization prescriptions themselves), with the use of (\ref{F17}),
under assumption that the corresponding relation between vertex and
self-energy diagrams is valid for all subgraphs and reduced graphs.
A crucial point is that in the right hand side of (\ref{F17})
one has reduced graphs with at least one photon line if
the initial graph has a photon line, because summation
in the right-hand side is over subgraphs $\gm$ such that
$l\overline{\in} \gm$.

The Ward identities that connect $N$-point and $N-1$-point Green
functions, with $N>3$ and $N=2$, are analogously proved.
Only diagrams without photon lines require special treatment.
These are the one-loop polarization of vacuum, vertex and box diagram.
However, the second one is zero, the third is convergent.
Thus, to complete consideration of the case $N=2$ it is sufficient
to prove that the one-loop polarization of vacuum is transverse.
A straightforward application of
general prescription (\ref{F17}) leads to a non-transverse expression.
Of course, it is possible
to adjust finite counterterms and arrive at a gauge-invariant result.
(Why is it bad to do this just for {\em one diagram} of the theory?)
However, following the style of differential renormalization,
one can modify the corresponding prescriptions by pulling out
differential operators in a manifestly gauge-invariant way.
Using manipulations \cite{HL} based on standard properties of Bessel
functions,
the one-loop polarization of vacuum can be written
for $x\neq 0$ as
\bea
\Pi_{\mu \nu} (x) \equiv
-e^2 \left( \frac{i m^2}{4\pi^2} \right)^2 \mbox{tr} \left\{
(m + i \pasl ) f(z) \gm_{\mu} (m - i \pasl ) f(z) \right\} \nn \\
= \frac{e^2 m^4}{12 \pi^4} (\pa_{\mu} \pa_{\nu} -g_{\mu\nu}) h(z),
\label{PLB}
\eea
where
\bea
f(z) = K_1 (z) /z, \\
\label{fz}
h(z) = K_1 (z) /z^2 +K_0 (z) K_1 (z) /z + K_0^2 (z) -K_1^2 (z),
\label{hz}
\eea
$z=m \sqrt{-x^2 +i 0}$, and $K_{0,1}$ modified Bessel functions.

Since the transverse structure is already explicit one can
remove divergences in the right-hand side of (\ref{PLB}) in an arbitrary
way. For example, it is possible to continue to pull out laplacians as
it was down in ref.~\cite{HL}.
However only the first term in the right-hand side of
(\ref{hz}) is UV-divergent (i.e. non-integrable in the vicinity
of the point $x=0$). In fact, it is proportional to the one-loop
scalar diagram so that one can also apply prescription
(\ref{prd}) with $\om=0$. Note that all four terms in the
right-hand side of (\ref{hz}) admit a natural diagrammatical
interpretation and have simple expressions in momentum space,
because $K_0(z)$ is obtained from $K_1/z$ (i.e. scalar propagator)
by the operator $-m \frac{\dd}{\dd m}$, and $K_1$ just by multiplication
by $m^2 x^2$ (i.e. $-m^2 \Box_q$ in momentum space).

To conclude the section we note that in other theories with Abelian
gauge symmetry the situation is quite similar, with unessential
additional considerations. For example, in scalar electrodynamics,
one should take into account massive tadpoles (which are zero in QED).
The tadpoles are not still covered by prescriptions
(\ref{F17})--(\ref{F18}). However, in the next section, we shall
arrive at general prescriptions, using the momentum space language.

\section{Differential renormalization \newline in momentum space}

\subsection{General renormalization prescriptions}

Let us now turn to momentum space where basic physical quantities
are calculated. First we observe that renormalization prescriptions
(\ref{F17})--(\ref{F18}) are trivially transformed into momentum space.
(We now distinguish external and internal vertices --- see comment~5
in the end of Sect.~2.)
Let $F_{\Gm} (\uq,\um) \equiv F_{\Gm}(q_1, \ldots, q_n, m_1,\ldots,m_L)$
be the Feynman integral given by
\bea
(2\pi)^4 \dl\left(\sum q_j\right)F_{\Gm}(\uq,\um) =
\int \dd x_{1} \ldots \dd x_n \exp\left(i\sum q_j x_j\right)
F_{\Gm} (\ux,\um) ,
\label{Fourier} \\
F_{\Gm} (\uq,\um) =
\int \dd k_1 \ldots \dd k_h \tilde{\Pi}_{\Gm} (\uq,\um),
\label{FI}
\eea
where $\uk\equiv k_1, \ldots k_h$ is a set of loop momenta of $\Gm$, and
$\tilde{\Pi}_{\Gm}$ is the product of propagators in momentum space
$\tilde{G}_l (q) = P_l (q, m) /(m_l^2-q^2-i0)$
associated with the given graph.
Then we get the following prescriptions:
\bea
R\, F^{j,l}_{\Gm}  = \frac{1}{j+1}
\left(\om /2 - \tilde{D} \right)\left (1-M^{\om-1}\right)
\; R'\, F^{j+1,l}_{\Gm} -
\frac{1}{j+1} \sum_{\gm \subset \Gm:\; l\overline{\in} \gm} R
\, {\cal C}_{\gm} F^{j+1,l}_{\Gm},
\label{F17m} \\
\Dl(\Gm) F^{j,l}_{\Gm} = R F^{j,l}_{\Gm} - R'\, F^{j,l}_{\Gm} ,
\label{Dlm} \\
\left(\om /2 - \tilde{D} \right) R\, F_{\Gm} =
\sum_{\gm \subseteq \Gm} {\cal C}_{\gm} R \Pi_{\Gm}
\equiv \sum_{\gm \subset \Gm}
R {\cal C}_{\gm} F_{\Gm }
+ {\cal C}_{\Gm} F_{\Gm } .
\label{F18m}
\eea

Thus the only distinction is that instead of operator $\SH$ we now have
dilatation operator (times 1/2)
\be
\tilde{D} = \frac{1}{2} \sum_{i} q_{i} \frac{\pa}{\pa q_{i}}
+ \frac{1}{2} \sum_{l} m_{l} \frac{\pa}{\pa m_{l}} .
\label{DQM}
\ee
By definition it acts now on integrands of Feynman integrals
{\em before integration} in loop momenta.

For example, for one-loop scalar Feynman integral with
general masses we have
\bea
R\; \int \dd k
\frac{\ln^j (m_1^2-k^2-i0)/\mu^2}{(m_1^2-k^2-i0)(m_2^2-(q-k)^2-i0)} \nn \\
= \frac{1}{j+1} \int \dd k
\frac{1}{2}\left( q \frac{\pa}{\pa q} + m_1 \frac{\pa}{\pa m_1}
+ m_2 \frac{\pa}{\pa m_2} \right)
\frac{\ln^{j+1} (m_1^2-k^2-i0)/\mu^2}{(m_1^2-k^2-i0)(m_2^2-(q-k)^2-i0)} .
\label{1lm1m2}
\eea
In particular, for $m_1=m_2=0$ and $j=0$, this reduces to
\be
R\; \int \frac{\dd k}{(-k^2-i0)(-(q-k)^2-i0)}
= - \int \dd k
\frac{q(q-k) \ln (-k^2-i0)/\mu^2}{(-k^2-i0)(-(q-k)^2-i0)^2} .
\label{1l00}
\ee

To calculate (\ref{1l00}) it is better not to differentiate
the integrand explicitly by operator $\frac{1}{2} q \frac{\pa}{\pa q}$.
Rather, it is worthwhile to introduce  analytic
regularization \cite{Speer}:
\be
RF(q) = \left. \int \dd^4 k \left(-\frac{\dd}{\dd \lm} \right)
\frac{1}{2} q \frac{\pa}{\pa q}
\frac{\mu^{2\lm}}{(-k^2-i0)^{1+\lm} (-(q-k)^2-i0)}
\right|_{\lm=0} .
\label{R1La}
\ee
When $\lm\neq 0$, we may use the following order: to calculate the integral,
differentiate in $q$, differentiate in $\lm$ and finally put $\lm=0$.
When calculating the integral one uses the four-dimensional one-loop
formula
\be
\int \dd^4 k \frac{1}{(-k^2-i0)^{\lm_1} (-(q-k)^2-i0)^{\lm_2}}
= i \pi^2 G(\lm_1,\lm_2) \frac{1}{(-q^2-i0)^{\lm_1+\lm_2-2}},
\label{1L}
\ee
where $G$ is the four-dimensional $G$-function
\be
G(\lm_1,\lm_2) =
\frac{\Gm(\lm_1+\lm_2-2)}{\Gm(\lm_1) \Gm(\lm_2)}
\frac{\Gm(2-\lm_1) \Gm(2-\lm_2)}{\Gm(4-\lm_1-\lm_2)}.
\label{4G}
\ee
In particular,
\be
G(1,1+\lm) = \frac{1}{\lm (1-\lm)} .
\label{4Ge}
\ee
Finally we have
\be
R \int \dd^4 k \frac{1}{(-k^2-i0) ((q-k)^2-i0)}
= i \pi^2 \left(1 - \ln (-q^2-i0)/\mu^2 \right) \, .
\label{R1L0}
\ee

For example~2 of Sect.~2 with a nest of divergent subgraphs,
as well as for other examples, momentum space
versions are quite similar, with product of propagators in the
Feynman integral in momentum space and operators $\tilde{D}$
(instead of $-\hat{S}$)
associated with external momenta and internal masses of corresponding
subgraphs.

Note that the arguments used to prove the coordinate space
prescriptions can be also translated in momentum space
language.\footnote{This was done
in ref.~\cite{Z} where more general prescriptions were formulated
for logarithmically divergent diagrams with simple topology
of subdivergences.} In fact, one starts with an incompletely
renormalized Feynman integral and writes down the formula of
(strictly four-dimensional) integration by parts.
Then the coordinate space procedure of extension of the product of
distributions to the whole space looks, in momentum space, as dropping
surface terms in this formula (which are polynomials in external momenta
and correspond to counterterms). To resolve the structure of (generally
overlapping) divergences one here uses sectors in momentum space.

Let us now remember that we considered our prescriptions
(\ref{F17})--(\ref{F18}) and their momentum space versions
(\ref{F17m})--(\ref{F18m}) only
for graphs that do not contain tadpoles and that do not lead
to tadpoles when inserting associated counterterms.
It turns out that the momentum space prescriptions already have
a desired form that is applicable for general graphs.
In particular, for the tadpole graph shown in Fig.~1a,
\setlength{\unitlength}{1mm}
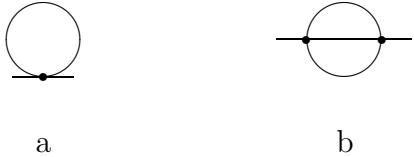
\begin{figure}[htb]
\begin{center}
\begin{picture}(150,28)(-15,0)
\put (40,20) {\circle{10}}
\put (40,15) {\circle*{1}}
\put (36,15) {\line(1,0){8}}
\put (39,5) {a}
\put (80,20) {\circle{10}}
\put (75,20) {\circle*{1}}
\put (85,20) {\circle*{1}}
\put (71,20) {\line(1,0){18}}
\put (79,5) {b}
\end{picture}
\end{center}
\caption{(a) tadpole; (b) sun-set diagram}
\end{figure}
the recipe (\ref{F17m}) (that is obtained using the above arguments
based on integration by parts in momentum space)
gives
\bea
R\; \int \frac{\dd k}{m^2-k^2-i0}
= \int \dd k \left(1- \frac{1}{2} m \frac{\pa}{\pa m}\right)
(1-M_m^1) \frac{\ln (m^2-k^2-i0)/\mu^2}{m^2-k^2-i0} \nn \\
=  \int \frac{\dd k}{(m^2-k^2-i0)^2}
\left\{ [m^2-k^2 \ln (1+m^2/(-k^2-i0))] \right. \nn \\
\left. - 2m^2 \ln (1+m^2/(-k^2-i0))
+(m^4/k^2) \ln (-k^2-i0)/\mu^2 \right\} .
\label{tadpole}
\eea
If we spoil the tadpole by $\ln (-k^2-i0)/\mu^2$, rather
than by $\ln (m^2-k^2-i0)/\mu^2$ we get a more simple result:
\be
R\; \int \frac{\dd k}{m^2-k^2-i0}
=  m^4 \int \frac{\dd k \ln (-k^2-i0)/\mu^2}{(-k^2-i0)(m^2-k^2-i0)^2} .
\label{tadpole1}
\ee

Let us now consider the sun-set diagram shown in Fig.~1b.
If we used coordinate space arguments and started from the incompletely
renormalized diagram
\[ R' \Pi_{\Gm} = (1+\Dl(12)+\Dl(23)+\Dl(31)) \Pi_{\Gm} \]
(where $\Dl(12)$,\ldots are counterterms associated with three
overlapping one-loop subgraphs)
and wanted to extend this functional from the space with deleted
origin, $x=0$, to the whole space, we would observe that
the one-loop counterterms still vanish once we have $x\neq 0$.
Thus we do not have enough space to perform renormalization
of the sun-set diagram (and other similar diagrams) in two steps
because extension to all $x$ requires simultaneous introduction
of the overall counterterm as well as counterterms for subgraphs.

To overcome this complication
and arrive at general prescriptions we used, in \cite{S2}, a trick
of ref.~\cite{L} based on Fourier transform with respect to the squares
of masses which were treated as squares of two-dimensional vectors.
A better solution of this problem is just to use momentum space
prescriptions (\ref{F17m})--(\ref{F18m}) that happen to be more flexible.
We state that these very prescriptions (\ref{F17m})--(\ref{F18m})
{\em are valid for arbitrary diagrams}. For example, in the case
of the sun-set diagram, we have
(for definiteness, we have chosen the first line to `spoil by a logarithm'
for renormalization of the whole graph)
\be
R\; F_{\Gm} =
(1 - \tilde{D}_{q,m}) (1-M_{q,m}^{1})  R'\, F^{1,1}_{\Gm}
-  c_{23} R\; \int \dd k \frac{\ln (m^2-k^2-i0)/\mu^2}{m^2-k^2-i0} .
\label{sun}
\ee
Here $F^{1,1}_{\Gm}$ differs from $F_{\Gm}$ by the additional
factor $\ln (m_1^2- p_1^2-i0)/\mu^2$ in the Feynman integral
($p_1$ is the momentum of this line),
the renormalized value for the tadpole (with a logarithm)
corresponding to the reduced graph $\Gm/\{23\}$ is given by
\be
R\; \int \dd k \frac{\ln (m^2-k^2-i0)/\mu^2}{m^2-k^2-i0}
= \frac{1}{2} \int \dd k
\left(\frac{1}{2} m \frac{\pa}{\pa m}-1\right) (1-M_m^1)
\frac{\ln^2 (m^2-k^2-i0)/\mu^2}{m^2-k^2-i0} ,
\label{tadpole_ln}
\ee
with $m=m_1$, and the constant $c_{23}$
is found from eq.~(\ref{F18m}), i.e. in our case it is proportional to
$\tilde{D} R F_{23}$
(in fact it does not depend on the renormalization scheme
--- compare \cite{S2}). Finally, counterterms for the given
graph spoiled by the logarithm (that contribute to $R'\, F^{1,1}_{\Gm}$),
are found from eq.~(\ref{1lm1m2}).

\subsection{VVA anomaly as an example}

Since differential renormalization is strictly four-dimensional
it looks preferable for application in each situation when dimensional
renormalization \cite{DIMREG}
meets difficulties, for instance, in theories with
chiral and super symmetries (see, e.g. \cite{appl1} where the initial
version of differential renormalization \cite{FJL} was successfully applied
in such cases, in particular for calculation of
anomalies). It is well-known that within dimensional
renormalization the origin of anomalies turns out to be an inconsistency
in definition of $\gm_5$ in dimensional regularization,
and strictly in four dimensions one does not have such problems at all.
Thus it is natural to ask where do the anomalies come from. As it was
demonstrated in \cite{appl1}, the anomalies within differential
renormalization appear because a system of
linear equations for finite counterterms to satisfy all Ward identities
turns out to overdetermined.

To see that the presented version of differential renormalization
is calculationally simple and
well suited for dealing with chiral problems let us once again
consider the calculation of the ABJ triangle anomaly \cite{ABJ}
as an example.
Let $T_{\al\beta\rho} (q,p) =
S_{\al\beta\rho} (q,p)+ S_{\beta\al\rho} (-q-p,p)$ be the sum
of two triangle diagrams contributing to one-loop Green function
of the axial current $J_{5\rho}=\bar{\psi}\gm_{\rho}\gm_5 \psi$, and
let $p$ correspond to the axial current and $q,-q-p$ to other
two external lines. In accordance with general prescriptions
(namely eq.~(\ref{prd}) translated into momentum space language),
the differentially renormalized quantities are given by
\bea
R_j S_{\al\beta\rho} (q,p) =
- i \int \frac{\dd k}{(2\pi)^4} (\tilde{D}_{q,p,m}-1/2)(1-M^{0}_{q,p,m})
\ln(m^2-p_j^2-i0)/\mu^2 \nn \\
\times \mbox{tr}\left\{
\gm_{\rho}\gm_5 \frac{1}{m-\ksl} \gm_{\al} \frac{1}{m-\ksl-\qsl}
\gm_{\beta} \frac{1}{m-\ksl+\pesl}
\right\} ,
\label{tri}
\eea
where $m^2-i0$ prescriptions are omitted for brevity.
Here $j=AV,VV,VA$ and $p_j$ can be chosen as the
momentum of the corresponding line (between vector or axial vertices).
According to remark in subsect.~2.2 one can use an
appropriate linear combination
$R =\sum_j \xi_j R_j$, and we will do this below.

Since we want to have Ward identities in both vector channels,
\be
q^{\al} R T_{\al\beta\rho} (q,p) =
(q_{\beta}+p_{\beta}) T_{\al\beta\rho} (q,p) =0 ,
\label{WIa}
\ee
we should introduce the logarithms symmetrically in AV and VA lines,
namely, choose $\xi_{AV}=\xi_{VA}$ in the above linear combination.

To calculate
$q^{\al} R_j T_{\al\beta\rho} (q,p)$
we use simple commutation relations (\ref{CR}) and
introduce auxiliary analytic regularization
(as in an example in subsect.~4.1).
by
$\ln(m^2-p_j^2-i0)/\mu^2 \to
-\frac{\dd}{\dd \lm} \frac{\mu^{2\lm}}{(m^2-p_j^2-i0)^{\lm}}$.
Then we observe that the action of the operator
$(\tilde{D}_{q,p,m}-1/2)(1-M^{0}_{q,p,m})$ reduces to
calculation of the finite part of the Laurent expansion in $\lm$
(actually the pole part turns out to be zero):
\bea
q^{\al} R_j S_{\al\beta\rho} (q,p) =
- i \int \frac{\dd k}{(2\pi)^4}
\frac{\mu^{2\lm}}{(m^2-p_j^2-i0)^{\lm}}
\nn \\
\left. \times q^{\al} \mbox{tr}\left\{
\gm_{\rho}\gm_5 \frac{1}{m-\ksl} \gm_{\al} \frac{1}{m-\ksl-\qsl}
\gm_{\beta} \frac{1}{m-\ksl+\pesl}\right\} \right|_{\lm=0} \, .
\label{tri1}
\eea

Now we apply momentum space version of (\ref{id}),
\be
\frac{1}{m-\pesl} \qsl \frac{1}{m-\pesl-\qsl}
=\frac{1}{m-\pesl-\qsl}  -  \frac{1}{m-\pesl} .
\label{idmom}
\ee
{}From coordinate space considerations, one sees that the result
should be polynomial of the second degree in $q,p,m$. Furthermore,
well-known formulae for traces of products of gamma matrices
show that one can put $m=0$. Then one uses relation
$\mbox{tr} \gm_{\mu} \gm_{\nu}\gm_{\al}\gm_{\beta}\gm_{5}
=4i\eps_{\mu\nu\al\beta}$, eqs.~(\ref{1L}), (\ref{4G}),
\be
\int \dd^4 k \frac{k^{\nu}}{(-k^2-i0)^{\lm_1} (-(q-k)^2-i0)^{\lm_2}}
= i \pi^2 G^{(1)}(\lm_1,\lm_2) \frac{q^{\nu}}{(-q^2-i0)^{\lm_1+\lm_2-2}},
\label{1Lk}
\ee
with
\be
G^{(1)}(\lm_1,\lm_2) =
\frac{\Gm(\lm_1+\lm_2-2)}{\Gm(\lm_1) \Gm(\lm_2)}
\frac{\Gm(3-\lm_1) \Gm(2-\lm_2)}{\Gm(5-\lm_1-\lm_2)} ,
\label{4G1}
\ee
and the value (\ref{4Ge}) as well as
\be
\left.\left\{ G(1+\lm,1) - 2 G^{(1)} (1+\lm,1) \right\}\right|_{\lm=0}
=1/2, \;\;\;
\left.\left\{ G^{(1)} (2,\lm) \right\}\right|_{\lm=0} = 1/2 .
\label{pava}
\ee
The final result is
\be
q^{\al} R_{AV} T_{\al\beta\rho} =
q^{\al} R_{VA} T_{\al\beta\rho} = \frac{i}{8\pi^2}, \;\;\;
q^{\al} R_{VV} T_{\al\beta\rho} = \frac{i}{4\pi^2}
\label{qRi}
\ee
so that we have Ward identities in the vector channels (\ref{WIa})
for $R=R_{AV}+R_{VA}-R_{VV}$.

Within this choice of renormalization one can perform calculation of
$p^{\rho} R S_{\al\beta\rho} (q,p)$ using
\be
\frac{1}{m-\pesl} \qsl \gm_5 \frac{1}{m-\pesl-\qsl}
=\gm_5 \frac{1}{m-\pesl-\qsl}  -  \frac{1}{m-\pesl} \gm_5
- 2m \frac{1}{m-\pesl}  \gm_5 \frac{1}{m-\pesl-\qsl},
\label{idmom5}
\ee
instead of (\ref{idmom}), and the same technique as before, in particular
formulae (\ref{1L}), (\ref{4G}), (\ref{1Lk}), (\ref{4G1}),
(\ref{pava}), with the result
\be
p^{\rho} R T_{\al\beta\rho} =
2m i R T_{5\al\beta}
- \frac{i}{2\pi^2} \eps_{\al\beta\mu\nu} p^{\mu} q^{\nu},
\label{pRi}
\ee
where $T_{5\al\beta}$ is one-loop contribution to the Green function
of the pseudovector current $i \bar{\psi}\gm_{\rho}\gm_5 \psi$.
This leads to the well-known value of the VVA anomaly.

\section{Conclusion}

The prescriptions of differential renormalization scheme presented above
are applicable for arbitrary diagrams and look more simple as compared
with previous versions. Let us stress that the main features of differential
renormalization are pulling out certain differential operators and
introducing a logarithmic dependence into the diagrams involved.
One can define it either in coordinate or momentum space
(although momentum space turns out to be more flexible in some
situations).

Since the mechanism of proving Ward identities in QED at the diagrammatical
level is very transparent, it was possible to do this automatically,
to all orders. Of course, the diagrammatical realization of non-Abelian
gauge symmetries is rather non-trivial. One may certainly hope that
the presented version of differential renormalization can be useful
in treating non-Abelian gauge symmetries combined with super and
chiral symmetries strictly in four dimensions
at least in lower orders of perturbation theory.
Since the differential
operators that are pulled out during renormalization have very simple
commutation relations with multiplication by momenta,
the problem here is to control the logarithms that are generated
by renormalization.

Finally we note that the presented version of differential renormalization
is naturally supplied with strictly four-dimensional methods of calculation
of Feynman integrals, in particular integration by parts \cite{S3}
which is a four-dimensional analogue of the method of integration by parts
within dimensional regularization \cite{IBP}
and is itself based on the differential renormalization and
its infrared analogue.

{\em Acknowledgments.}
I am grateful to K.G.~Chetyrkin and D.Z. Freedman for valuable discussions.

\end{document}